$s+id pairing was criticized. However, the author had overlooked a
\documentstyle[preprint,revtex]{aps}
\begin{document}
\draft
\begin{title}
\center{Weak coupling model for s*- and d-wave superconductivity}
\end{title}
\author{D.van der Marel}
\begin{instit}
Laboratory of Solid State Physics, Materials Science Center\\
University of Groningen, Nijenborgh 4, 9747 AG Groningen
\end{instit}
\begin{abstract}
The phase diagram of the unconstrained $t-J$ model is calculated
using the random phase approximation. It is found that the extended
$s$ and the $d_{x^2-y^2}$-channels are {\em not} degenerate near half filling.
Extended $s$-pairing with a low $T_c$ occurs only for a band containing less
then
0.4 electrons or holes per unit cell,
whereas in a large region around half-filling $d$-wave pairing is the only
stable superconducting solution. At half filling superconductivity is
suppressed
due to the formation of the anti-ferromagnetic Mott-Hubbard insulating state.
By extending the analysis to the unconstrained $t-t'-J$ model, it is proven
that,
if a Fermi surface is assumed similar to the one that is known
to exist in cuprous oxide superconductors, the highest superconducting $T_c$ is
reached for about $0.7$ electron per site, whereas
the anti-ferromagnetic solution still occurs for $1$ electron per site.
It is shown, that the maximum $d$-wave superconducting mean field transition
temperature is half the maximum value that the Ne\`el temperature
can have in the Mott-insulating state.
\end{abstract}
\pagebreak
In spite of a huge experimental and theoretical effort to
understand the superconductivity in the cuprous oxide
high T$_c$ superconductors\cite{bednorz,wu}, a mechanism for superconductivity
has not yet been firmly established. A lot of attention has been
devoted to obtain a model for superconductivity starting from
the Hubbbard hamiltonian, however, there is a growing suspicion
that the positive U Hubbard model alone can not give rise to
superconductivity\cite{sudbo}. A different approach has been to treat
the electronic degrees of freedom and 'spin fluctuations effectively as
separate channels\cite{scalapino,monthoux2}, leading to a
retarded electron-electron interaction mediated by
spin-fluctuations. The latter model has proven to be
more successful in the sense of providing a possible mechanism
of superconductivity.
\\
Here I will discuss pairing using an exchange-only Hamiltonian
\begin{equation}
\begin{array}{lll}
H &=& \sum_{k,\sigma} (\xi_k-\mu)c_{k\sigma}^{\dagger}c_{k\sigma} \\
  &+& 2\sum_{Q}\sum_{k,q}J(Q) \left[ \vec{S}_{k,k+Q} \cdot \vec{S}_{q,q-Q} -
 \frac{1}{4}\sum_{\sigma\sigma'}c_{k+Q\sigma}^{\dagger}c_{q\sigma'}^{\dagger}
            c_{q+Q\sigma'}c_{k\sigma} \right]
\label{eq:jss}
\end{array}
\end{equation}
The $t-J$ model is studied here without the usual constraint on double
occupancy of the same site as a model in its own right.
In the real-space representation $J$ couples nearest
neighbouring sites on a square lattice. Hence the exchange part is of the form
$\frac{1}{2}(J_0\sum_{R,\delta}\vec{S}_{R}\cdot\vec{S}_{R+\delta}
-\frac{1}{4}n_Rn_{R+\delta})$
where $\delta$ runs over the four nearest neighbour sites,
and the factor $1/2$ compensates double counting of
the interactions in the summation over $R$. If one considers only
two neighbouring spins the energy of the triplet state is
$0$, and of the singlet it is $-J_0$, hence the sign convention
is such, that $J_0>0$ couples neighbouring spins anti-ferromagnetically.
\\
For the dispersion of the band
I will consider a nearest neighbour and a next nearest neighbour hopping
term, but apart from that, it is the same hamiltonian as was used
in the papers by Baskaran, Zou and Anderson (BZA)\cite{bza1,bza2},
Emery\cite{emery}
and by Kotliar
\cite{kotliar}. Although usually such a Hamiltonian is thought to be
derived from the Hubbard model by means of a Gutzwiller projection,
(which also changes the character of the fermion
operators, by projecting out double occupancy of the same sites) it
should be pointed out, that in the cuprous oxide systems
this term may also have a different microscopic origin. As the
actual bandstructure
in these systems is experimentally known to be better described
by the three band model of Zaanen, Sawatzky and Allen\cite{zsa},
(which is again a simplified version of the real valence band structure
involving 6 oxygen 2p bands and 5 cupper 3d-bands for the occupied
states, as well as unoccupied 3s and 3p states)
a transformation to a single band hamiltonian will in principle
generate both an effective Hubbard $U$ and an
intersite $J$\cite{zr,es,jef}. Examples of such transformations can be found
in the work by Emery \cite{emery}, and by Jansen \cite{jansen}.
However, also other, more complicated types of interactions are generated when
making transformations of this type, notably the correlated hopping term (with
six operators) which, as has been shown by Hirsch, promotes
superconductivity of hole-carriers\cite{hirsch}. The interaction
considered by Jansen as well as the correlated hopping term treated by Hirsch,
effectively provide an {\em on-site} attraction,
which, when considered on its own, promotes pairing in the (non-extended)
s-wave channel.
In this paper I will assume that the net on-site interaction is repulsive,
which, as will be discussed below, tends to suppress superconductivity by
stabilizing
the anti-ferromagnetic solution.\\
Monthoux and Pines\cite{monthoux2} have considered the $t-t'$ bandstructure
together with
an interaction of the form $g(q)\vec{s}(q)\cdot\vec{S}(-q)$, where $\vec{s}(q)$
represents
the valence electron spin-operator, and $\vec{S}(-q)$ is a separate
spin-fluctuation operator, the properties of which are determined by the
spin-susceptibility.
The transport and superconducting properties are then
calculated from strong-coupling theory using empirical values for the
spin-susceptibility
and $g(q)$. It has been shown by Monthoux {\em et al.} that the effective
electron-electron
Kernel arising from such a coupling becomes indeed
a spin-dependent interaction\cite{monthoux1}, which could in principle be
written
as a frequency dependent version of Eq. \ref{eq:jss} \cite{footnote}.
In the same paper a weak coupling analysis
of such a frequency dependent Kernel was given. In the present paper
the problem is further simplified by disregarding a possible frequency
dependency of $J(Q)$,
which, depending on the details of the microscopic origine of Eq. \ref{eq:jss},
may still be a
justifyable approximation. This allows us to explore the phase diagram in
somewhat more
detail without having too many parameters to consider.\\
Apart from these general considerations
I have no sound microscopic justification for using
this hamiltonian. The main motivation to use it comes from
the fact that, as I will show in this paper, it appears to do a
surprisingly good job as a phenomenological model
consistent with at least some of the experimentally known aspects
of superconductivity in these systems.  \\
BZA\cite{bza2} considered pairing of the $s^*$-type
near half filling, Emery considered $d_{x^2-y^2}$-pairing,
and Kotliar studied both $s^*$ and $d$-type pairing.
Below I will show, that the $s^*$-type pairing is not a stable
solution near half filling, and is dominated by pairing of the $d$-type.
As the latter again tends to be unstable with respect to the
anti-ferromagnetic Mott-Hubbard insulating state at half filling,
superconductivity can only exist sufficiently
far away from this region. As the optimal $T_c$ would
have been reached at half filling for a symmetrical band, this
would lead to the conclusion that superconductivity is only
a marginal effect in such a system. However, the high $T_c$ cuprates
do not have a symmetrical band, and the Fermi surface is known to be
distorted from the perfect square that arises from considering
only nearest neighbour hopping. This actually comes to rescue:
As a function of band-filling it pulls apart the regions, where
anti-ferromagnetism and high $T_c$ are optimal,
{\em without} having a noticable effect on the superconducting
or anti-ferromagnetic transition temperatures. Three important
trends emerge from this analysis:\\
(1) Given the distorted shape of the Fermi surface as it is known
 to occur in the cuprates, hole doping
 gives rise to higher $T_c$'s than electron doping. \\
(2) There exists a universal relation between the
 highest Ne\`el temperature found in the phase diagram and the highest
 possible mean field superconducting $T_c$, with $T_N/T_c\approx 2$.
 That a relation of this kind should exist was already pointed out
 by Anderson\cite{philisalwaysright} shortly after the discovery by
 Bednorz and M\"uller.\\
(3) This implies that with reasonable values for the intersite exchange
 interaction, providing the correct Ne\`el temperature, one automatically
 obtains values for the superconducting transition temperature
 which are (even though they are larger than the experimental values
 in the cuprates) definitely in the right ball-park. \\
 \\
The $k$-space representation of the exhange interaction
is of the form of Eq.\ref{eq:jss} with
 \begin{equation}
 J(Q)=\frac{1}{2} J_0(\cos(Q_xa)+\cos(Q_ya))
 \end{equation}
This type of interaction favours anti-ferromagnetism if
$J_0>0$, which becomes especially stable
if the band is half filled. The antiferromagnetic alignment
of nearest neighbours gives rise to a spin-dependent effective
field, which is periodic with the wave vector $(\pi/a,\pi/a)$.
\\
Let us now turn to the superconducting gap equations.
If the interaction potential $V_{kq}$ is of the form
 \begin{equation}
  H^i=-\sum_{k,q}V_{k,q}
  c_{k\uparrow}^{\dagger}c_{-k\downarrow}^{\dagger}
  c_{-q\downarrow}c_{q\uparrow}
 \end{equation}
the BCS gap equation is \cite{BCS}
\begin{equation}
 \Delta_k=\sum_q\frac{\Delta_q V_{k,q}}{2E_q}\tanh(\frac{E_q}{2k_BT})
\end{equation}
where $E_q\equiv\sqrt{\epsilon_q^2+\Delta_q^2}$ as usual. If we can
make the assumption, that the main contribution leading to superconductivity
comes from the $J$-term, we see that the interaction
entering the gap equations is
 \begin{equation}
  V_{k,q}=2J(k-q)+2J(k+q)
 \end{equation}
With this substitution we obtain
 \begin{equation}
 \Delta_k= J_0\left[ \Psi_x^+\cos{k_xa} + \Psi_y^+\cos{k_ya} \right]
 \end{equation}
where I introduced the dimensionless pairing amplitudes
 \begin{equation}
 \begin{array}{lll}
 \Psi_{i}^+&\equiv&
  \sum_{q}E_q^{-1}\Delta_q\cos{q_ia} \tanh \left( \frac{E_q}{2k_B T} \right)\\
 \end{array}
 \end{equation}
As there are two possible order parameters $\Psi_i^+$, we have here
two coupled equations, which can be easily disentangled
with the help of symmetry selection rules. I will do this for the
case where the
superconductor has a four-fold rotation axis. In that case $\Delta$
is either an odd or an even function of $k$. In the former
case, which corresponds to $d_{xy}$ symmetry, pairing amplitudes
of the form $\Psi_i^-= \sum_{q}E_q^{-1}\Delta_q\sin{q_ia}$ have
a finite amplitude, whereas $\Psi_x^+$ and $\Psi_y^+$ are zero.
The hamiltonian considered here does not couple to the $d_{xy}$
pairing-channel. If $\Delta$ is even, we have to
consider two possibilities: Either $\Psi_x^+=\Psi_y^+$ leading
to an extended s-wave gap, or $\Psi_x^+=-\Psi_y^+$ leading to
a $d_{x^2-y^2}$ symmetry of the gap function.
The gap function corresponding to these two cases is
 \begin{equation}
 \Delta_k=\frac{1}{2}\Delta_0 {[\cos{k_xa} \pm \cos{k_ya}]}
 \end{equation}
where the plus and minus sign correspond to the $s^*$- and
$d_{x^2-y^2}$-wave types of pairing respectively,
and the gap equation becomes

 \begin{equation}
 \frac{2}{J_0}=\sum_q E_q^{-1}
       {[\cos{q_xa} \pm \cos{q_ya}]}^2 \tanh(\frac{E_q}{2k_BT})
 \label{eq:gap}
 \end{equation}

This equation was also obtained by Kotliar \cite{kotliar}. In his analysis
the constraint of no double occupancy of the same site was taken
into account in an appriximate way, by having $t$ proportional to doping of
the half filled band. At half filling one then effectively has $t=0$, for
which case, as was shown by Kotliar, the summations on the right hand
side of this  Eq. \ref{eq:gap} are identical for the two
types of pairing. As a result he obtained a degeneracy between the
$d$-and $s$-ordered state at half filling, leading to the
conclusion that a pairing of type $s+id$ could occur. For any
finite value of $t$ this degeneracy is however lifted.
In the mean time a variety of numerical and theoretical techniques have
been applied to the $t-J$ and related models,
from which a tendency toward $d_{x^2-y^2}$-pairing has been found
near half filling\cite{gros,kotliar2,grilli,grilli2,dagotto}.
It is easy to show, that for the $s^*$-type pairing at half filling
of a symmetric band, $J_0$ has to exceed a critical value. Let us
assume that $\epsilon=(W/4)\cdot(\cos(k_xa)+\cos(k_ya))$. For $T=0$
the gap equation becomes
 \begin{equation}
  J_0^{-1}\sqrt{W^2/4+\Delta_0^2} =
       \sum_q \left|\cos{q_xa} + \cos{q_ya}\right| = 0.811
 \end{equation}
hence the cricial value of $J_0$ for $s^*$-pairing is at half filling
$J_0^c=0.62W$. The reason for the appearance of a critical value is, that
at half filling the $s^*$-type gap
is exactly zero for all $k$ at the Fermi surface. Only by mixing
in states away from the Fermi level, superconductivity of this type
may occur, which requires a minimum coupling strength. The $d$-channel
is much more effective in this sense, as $\Delta_k$ is finite at
the Fermi surface except for the node-points. As a result in the
$d$-channel we have $J_0^c=0$. \\
The ground state energy relative to the normal state
can now be determined by realizing that
it is the expectation value of the reduced hamiltonian
minus the non-interacting part, which is \cite{tinkham}

\begin{equation}
\sum_k\left(|\epsilon_k|-E_k+\frac{\Delta_k^2}{2E_k}\right)
\end{equation}

where the first
two terms represent the energy gained by redistributing the
electrons over $k$-space in the correlated wavefunction, whereas the
third term compensates double counting of the interaction.
In principle one has to solve the gap equation together
with a constraint on electron occupation number\cite{mu1,mu2,mu3}, however
the corrections to the free energy are of the order $(\Delta_0/E_F)^2$,
\cite{kirchberg} which is small for the parameters that
we will consider.
\\
I still need to specify the electron dispersion relation before
we can solve the
gap equations. If one considers a tight-binding model with
a single orbital per site, with only hopping between nearest and
next nearest neighbours, the single particle energies are
\begin{equation}
\epsilon_k=-2t\left(\cos(k_xa)+\cos(k_ya)\right)
   -4t'\cos(k_xa)\cdot\cos(k_ya) - E_F
\end{equation}
The $t'$-term is due to next-nearest neighbour hopping. Let me briefly
discuss some of the properties of such a band. If $t'=0$ at half filling
of the band, such a dispersion relation has the remarkable property
that the Fermi surface forms a perfect square, with a diverging
effective mass over the entire Fermi surface. In
practice this situation will never occur, as there will always be
some finite coupling between next nearest neighbours. This causes
a bulging of the Fermi surface, as is shown
in Fig. \ref{fig:fermi}, which eventually transforms into a rotated
Fermi surface if $|t'| \gg |t|$. The
shape obtained for $t'=-0.7t$
is very close to what has been calculated with the
local density approximation for {\em e.g.}
La$_2$CuO$_4$ and YBa$_2$Cu$_3$O$_7$ \cite{pickett,shen}. A significant
change also occurs in the density of states (DOS) at the
Fermi energy, which
is displayed in Fig. \ref{fig:nosdos} as function of the
number of electrons per unit cell. This
somewhat unusual representation of the DOS
is useful in the discussion below, where we
compare ground state energies of various types of ordering
at a given electron density. We see, that as $t'$
is increased, the DOS becomes
a-symmetric, and the maximum is shifted to the left side of the
point where the band is half filled. Of course the
direction in which this occurs is dictated by the sign of $t'$. With
$t'<0$ we mimic the situation encountered in
the $CuO_2$-planes of the high $T_c$ cuprates. \\
In Fig. \ref{efree:a} numerical calculations of the free energy are shown
as a function of occupation number for $J_0/W=0.6$, where $W=8t$ corresponds to
the bandwidth if $t'=0$. For the sake of completeness also the free
energy of the anti-ferromagnetically ordered state is included. This
was calculated from the same hamiltonian. To stay in the same spirit
as for the superconducting solutions, the random phase approximation was
used. Hence the free energy was minimized together with a constraint on
the electron occupation number, anticipating a finite expectation value of
$<c_{k+Q\uparrow}^{\dagger}c_{k\uparrow}>
=-<c_{-k\downarrow}^{\dagger}c_{-k-Q\downarrow}>$ at the point
$Q=(\pi/a,\pi/a)$. We notice
that the anti-ferromagnetically ordered state at half filling is always
more stable than the metallic state. However, for small
values of $J_0$ the $d$-wave paired superconducting state is still more
favourable. This is a consequence of our choice of model Hamiltonian,
which is perhaps somewhat pathological near half filling: Physically
the exchange terms should arise from a strong repulsive interaction
between electrons making a virtual transition to the same orbital of
{\em e.g.} a transition metal atom. On the one hand this
leads to exchange coupling between spins on neighbouring orbitals, while
on the other hand it causes the opening of a Mott-Hubbard gap,
which is much larger than the anti-ferromagnetic gap. This would
strongly stabilize the anti-ferromagnetic solution.
Tempting as it may be to add an on-site repulsion at this point as an
additional model parameter, I will not do so: It
has become clear in recent years, that a large
repulsive $U$ gives rise to very strong correlation effects, at and
near half filling, which can not be properly treated with the random
phase approximations made in this paper\cite{michael,nagaosa,bobl}.
For this reason, and also because fluctuations are neglected
with the latter approximations, the
present analysis is insufficient
close to half filling. For higher doping
it could have some relevance to the mechanism of
superconductivity. It is important to add in this context, that
the symmetry of an additional on-site interaction is such, that
it cancels out in the gap equation for the $s^*$ and $d$-channels.
Hence an on-site $U$ does not affect the gap-function or the free
energy for these types of superconductivity.
\\
Although from a Maxwell construction one is lead to the conclusion
that phase separation should occur in $s^*$- and $d$-ordered regions,
this is strongly suppressed if the long range Coulomb interaction is
taken into account.\cite{phasesep1} Although the Coulomb term is not
included explicitly in the Hamiltonian, the presence of such a
term is assumed implicitly by imposing the constraint that the
electronic density is macroscopically conserved.
As was stressed by Emery, Kivelson and Lin\cite{emery2,emery3},
who studied the $t$-$J$ model
together with the constraint on no doubly occupied sites, 'the holes are
often donated by oxygen atoms which are quite mobile ... ', providing
a physical mechanism for screening of the long range Coulomb term. Putikka,
Luchini and Rice\cite{putikka}
provided numerical evidence that, in the absence of a long range
Coulomb force, phase separation occurs for $J/t > 3.8$
as $n \rightarrow 0$ and $J/t > 1.2$ near half filling. The present
analysis does not lead to phase separation if
the long range Coulomb interaction is taken into
account.\cite{phasesep2}
\\
The phase diagram is displayed in Fig. \ref{phase}.
Due to electron-hole symmetry in
this case, the diagram is symmetric around half occupation
of the band. Roughly speaking $s^*$-pairing is favoured far away
from half filling of the band, whereas $d$-wave pairing becomes the
most stable solution near half filling.
For $J_0<0.3W$ there are regions of no superconductivity, which
broaden upon decreasing $J_0$, and completely cover the horizonal axis for
precisely $J_0=0$. This tendency towards $d$-wave pairing near
half filling was also obtained by Littlewood\cite{littlewood} for the charge
transfer model\cite{zsa}, again using a weak coupling treatment. In these
calculations an inter-site exchange interaction is not introduced explicitly,
and
can only result inderectly from the repulsive on- and inter-site interactions
which are taken into account in the model. \\
We see, that $J_0>0.7W$ is required to find an
antiferromagnetic phase near half filling. As can be seen from the
free energies versus doping, the phase boundaries between $s^*$ and
$d$, and between $d$ and $AF$, correspond to a discontinuous change
from one type of ordering to the other. For the $s^*$-$d$ boundary
this discontinuity will probably be softened without loosing the
superconductivity by the occurrance of an intermediate state of
mixed $s+id$ character, as was
proposed by Kotliar at precise half filling and $t=0$. The phase
boundary between $d$ and $AF$ is different in this respect. As
both the anti-ferromagnetic and superconducting correlations occur
in the same band of electrons, they will tend to suppress
each other. Because finite anti-ferromagnetic correlations will occur
on the superconducting side of the phase boundary and vice versa, at
the boundary $T_c$ and $T_N$ should come out to be zero if such corrections
are taken into account. This requires a treatment of the model hamiltonian
which goes beyond the level of random phase approximations
made in this paper.
The fact, that the $d$-paired and anti-ferromagnetic
solutions both have their optimum at half filling, is rather worrying,
as in a real solid the anti-ferromagnetic
solution will in practice turn out to be the more stable one, due to
the opening of a Mott-Hubbard gap.
\\
Fortunately nature does provide us with a way to make a
separation in parameter space between the
anti-ferromagnetic and superconducting states.
As already pointed out above, in practice there will always be
a finite value of $t'$. From Fig.
\ref{fig:nosdos} we see, that in this case the maximum value of
the DOS does {\em not} occur at half filling of the band. A
well-known result from BCS theory is, that a high DOS at the Fermi
level enhances $T_c$. If on the other hand precisely 50 $\%$ of
the states is occupied, the opening of an anti-ferromagnetic gap
causes a downward shift of all occupied levels in the reduced
Brillouin-zone, which is the reason why the anti-ferromagnetic
solution is best stabilized at precise half filling. This
effect is demonstrated in Fig. \ref{efree:b}, where we see that
indeed the lowest free energy of the $d$-paired state occurs now
at $35 \%$ filling, whereas the anti-ferromagnetic solution is
still at half filling. We also notice from this plot, that
this a-symmetry implies that the highest $T_c$'s of a $d$-paired
superconductor are to be expected on the left ('hole-doped')
side of half-filling. Lower $T_c$'s occur on the right side.
\\
Let us now consider the $\Delta/T_c$-ratio following from the gap equation.
Within the context of BCS theory we have $\Delta_0(T)=0$ at $T_c$,
so that $T_c$ follows from
 \begin{equation}
 \frac{2}{J_0}=\sum_q \epsilon_q^{-1}\tanh(\frac{\epsilon_q}{2k_BT_c})
       {[\cos{q_xa} \pm \cos{q_ya}]}^2
 \end{equation}
where the $\pm$ sign refers again to the two symmetries of pairing. This
equation can be easily solved numerically. The result is, that for extended
s-wave pairing the ratio $2\Delta_0/k_BT_c$ is 6.5, whereas for
d-wave pairing it rises gradually from 4 if $J_0 \ll W$,
up to 6.5 in the limit where $J_0 \gg W$. This is not sensitive to the
value of the parameter $t'$. We should keep in mind here, that
$\Delta_0$ is the maximum value reached by $\Delta(k)$
(respectively at the $(\pi,0)$- and $(\pi,\pi)$-point for $d$- and
$s^*$-pairing). \\
Finally it is interesting to look how the mean field estimate of
$T_c$ depends on the
coupling strength $J_0/W$. In Fig. \ref{fig:tcj0} $T_c^{MF}/W$ is
displayed as a function of $J_0/W$ for the d-wave channel. First
of all we notice, that for $J_0 > W/4$ the value of $T_c^{MF}$ is
about $J_0/4$. For $J_0/W << 1$ this crosses over to a quadratic dependency
$T_c^{MF}=4J_0^2/W$. For
comparison a similar curve is displayed for conventional s-wave pairing,
using the negative U Hubbard model in a band with a square DOS. We notice that
the mean field transition temperature with the latter model
becomes $T_c^{MF}=|U|/4$ for large $|U|$ (which is actually outside the
range of validity of the BCS weak coupling approach \cite{micnas,localized}),
and has the familiar BCS-like $\exp{(-W/|U|)}$ behaviour for small $U$.
The $T_c$ for the extended s-wave pairing lies again below the negative
$U$ curve, and is only finite above a threshold value of $J_0$ as discussed
above.\\
A consequence of this is, that the model hamiltonian proposed here
leads to quite reasonable values of the transition temperature, {\em which
are relatively insensitive to the value assumed for the bandwidth}. If we
assume that for example $J_0=0.1 eV$, we would find that the
Ne\`el temperature in the Mott-insulating
state can not exceed the mean field value for $Z=4$ interacting
neighbours $T_N^{MF}= ZS(S+1)J_0/6=580 K$. If we assume that
the bandwidth is smaller than about 0.5 eV, we obtain from
the BCS gap equation that the d-wave transition temperature
can not exceed the mean field value $T_c^{MF} = 290 K$. This demonstrates that
the optimal Ne\`el temperature and
the optimal superconducting transition temperature
have a ratio of about $2$ for $J_0/W$ of the order 0.2 to 1. Keeping
in mind, that with the mean field approach we over-estimate {\em both}
$T_N$ and $T_c$, I expect that the ratio between the two should remain
relatively
intact if corrections beyond the mean field approximation are included. In
the limit where $J_0/W$ is small, $T_c$ comes out smaller, although the
suppression
of the transition temperature goes much slower then for conventional
$s$-wave superconductivity. For example if the bandwidth is 1 eV we
obtain $T_c=137$ K, and with $W=2$ eV we find that $T_c=74$ K.\\
Finally it is possible now to draw a phase diagram in the temperature versus
doping plane. Let us choose $J_0=0.1 eV$, which gives approximately the
correct value for $T_N$, and $t'/t=-0.7$ which gives approximately the right
Fermi surface. To stay in the regime where $T_N\approx 2 T_c$, let
us assume $8t=W=0.5 eV$. The latter parameter is rather small
compared to the 2 to 3 eV of the bare copper-oxygen p$_x$,
p$_y$, d$_{x^2-y^2}$ anti-bonding band\cite{pickett}, and
leads to a slight over-estimation of T$_c$. Although the estimated Ne\'el
temperature can be indicated at half filling of the phase-diagram, the
present analysis has no bearing on the region near half filling, which
was left open for that reason. The phase diagram, displayed in
Fig.\ref{fig:tcn}, perhaps somewhat optimistically gives values
of $T_c$ above 200 K, which sofar has not been found experimentally.
According to the rule-of-thumb
that $T_c$ scales with $T_N$\cite{philisalwaysright},
one has to look for systems with relatively high Ne\`el temperatures
in order to reach room temperature superconductivity. A number of factors will
lower T$_c$ below the mean field value given here. First of all,
anti-ferromagnetic correlations will occur near the Mott-insulating state,
which tend to suppress the superconducting order. Very strong
on-site spin correlations are known to exist due to the large on-site
$U$, but $d$ and $s^*$-pairs are insensitive to this interaction channel,
as can be seen from the gap equation. Second, the mean field
approach links $T_c$ directly to the energy scale of the pair-breaking,
which again leads to an overestimation of T$_c$. The reason for this, is that
the long range phase-coherence can be lost in a dephasing-transition, if
the phase fluctuations have a lower energy scale than the pair-breaking
energy. This requires a better knowledge of the phase fluctuation spectrum
of the $d$-wave superconducting state.\\
Using a weak coupling BCS treatment of the $t$-$t'$-$J$ model, I have shown
that there
exists a universal
ratio of 2 between the Ne\`el temperature at half filling and the optimal mean
field
superconducting
transition temperature. If a realistic shape of the Fermi surface is
taken, the optimal $T_c$ occurs for $0.7$ electron per site, while the
Mott-insulating
antiferromagnetic state occurs at half filling. With these parameters, $T_c$ is
shown to be
lower for electron doping than for hole doping.\\
{\em Acknowlegments} It is a pleasure to thank Z. X. Shen for a stimulating
discussion and communication of
some unpublished results, and J. Lorenzano and G. A. Sawatzky for
useful comments on the manuscript.
This work is part of the research program of the "Stichting voor
Fundamenteel Onderzoek der Materie", which is financially supported by the
"Nederlandse Organisatie voor Wetenschappelijk Onderzoek".

\figure{First Brillouin-zone of a square lattice, with the occupied states
indicated as the
 shaded area. The lozenge indicates the perfectly nested Fermi surface.
\label{fig:fermi}}
\figure{Density of states at the Fermi energy in units of $1/W$ as a function
of electron
 occupation number
\label{fig:nosdos}}
\figure{Free energy difference with the normal state of the $s^*$-wave (solid)
and
 $d$-wave(long dashed) superconducting state and of the anti-ferromagnetic
state (short
 dashed curve) with $J_0=0.6W$ and $t'=0$. Energies are in units of $W$.
\label{efree:a}}
\figure{Phase diagram in the $J_0$-$n$ plane, where $n$ is the number of
electrons per
 unit cell.
\label{phase}}
\figure{Free energy difference with the normal state of the $s^*$-wave (solid)
and
 $d$-wave(long dashed) superconducting state and of the anti-ferromagnetic
state (short
 dashed curve) with $J_0=0.6W$ and $t'=-0.7t$. Energies are in units of $W$.
\label{efree:b}}
\figure{Solid curve: $T_c/J_0$ calculated for the d-wave channel of the
exchange-only model
 with $t'=0$ and 1 electron per site. The same curve is obtained
 for $t'=0.7$ with 0.7 electron per site. Open
 lozenges: $T_c$ of the $s^*$-wave channel with the latter parameters. Dotted
curve:
 $T_c/|U|$ versus $|U|/W$ for the negative $U$ Hubbard model taking a square
DOS.
\label{fig:tcj0}}
\figure{Phase diagram in the temperature-density plane with the parameters
$J_0=0.1eV$, $W=8t=0.5eV$, and $t'/t=-0.7$. The curves are interrupted in
the part near the middle, where the present analysis is physically meaningless.
The mean field Ne\`el temperature at half filling, using the same value for
$J_0$ as in the metallic regime, is indicated as a clover.
\label{fig:tcn}}

\begin{references}
\bibitem{footnote} A small difference is the absence of
the spin-independent second term of Eq. \ref{eq:jss} in the expressions of
Monthoux. It
implies that in their analysis the triplet channel is repulsive, and the
interaction
in the $s^*$
and $d_{x^2-y^2}$ channels is scaled with a factor $3/4$ compared to our
expressions. Also
a repulsive interaction in the $d_{xy}$-channel emerges.
\bibitem{phasesep1} The charging energy per unit volume for a
small fraction of spherical
$d$-type droplets dispersed in an $s^*$-type medium is
$\frac{8}{15}\pi^2e^2(n_d-n)^2r^5\cdot N/V$, where
$N=\frac{n-n_s}{n_d-n_s}\frac{V}{4\pi r^3/3}$ is
the number of droplets, $V$ is the sample volume, $n_d$, and $n_s$ are the
densities of the $d$ and $s$-type ordered states, and $n$ is the average
density. For $n_d$ and $n_s$ of comparible magnitude we interprete $r$ as
the characteristic length scale on which phase separation occurs
with a charging energy per unit volume
$\frac{2\pi}{5}e^2(n_d-n)(n-n_s)r^2$, which has the correct limiting
behaviour for small $n_s$ or $n_d$ and interpolates between these two
cases. In a layered electron gas one can instead consider
a cilindrical shape, for which the charging energy density is
$\frac{\pi}{4}e^2(n_d-n)(n-n_s)r^2$. The $r^2$ scaling behaviour
tends to reduce $r$. A lower bound occurs due to
quantum size effects: Due to quantum confinement the
kinetic energy of each electron increases with an amount of about
$\frac{\hbar^2 \Delta k^2}{2m^*}$. For particles
moving in a 2D plane $\Delta k^2 \approx \frac{2\pi}{r^2}$, so that the
increase in energy per unit volume is
$\frac{2\pi\hbar^2}{m^*r^2}(n_d-n)(n-n_s)(n_d+n_s-n)(n_d-n_s)^{-2}$, which
again interpolates between the two limits where $n_d$ or $n_s$ is small.
\bibitem{phasesep2} The minimum energy due to the combined effect of the
Coulomb interaction and quantum confinement occurs at
a length scale $r^4=8a_0(n_d+n_s-n)(n_d-n_s)^{-2}$.
The increase in energy per unit volume is then
$\pi\mbox{Ry}(n_d-n)(n-n_s)(n_d-n_s)^{-1}\sqrt{(2a_0)^3(n_d+n_s-n)}$, where
Ry and $a_0$ are the effective Rydberg and Bohr radius respectively.
This should be smaller than the gain in free energy due to phase separation.
For the parameters considered in this text, the latter is at least two
orders of magnitude smaller than the former, implying that phase separation
is strongly suppressed.
\end{references}
\end{document}